\documentclass[12pt]{article}
\usepackage{graphicx}

\begin{document}

\begin{center}
{\bf Black hole emission of vector particles in (1+1) dimensions} \\
\vspace{5mm} S. I. Kruglov
\footnote{E-mail: serguei.krouglov@utoronto.ca}

\vspace{3mm}
\textit{Department of Chemical and Physical Sciences, University of Toronto,\\
3359 Mississauga Road North, Mississauga, Ontario L5L 1C6, Canada} \\
\vspace{5mm}
\end{center}

\begin{abstract}
We investigate the radiation of spin-1 particles by black holes in (1+1) dimensions within the Proca equation. The process is considered as quantum tunnelling of bosons through an event horizon. It is shown
that the emission temperature for the Schwarzschild background geometry is the same as the Hawking temperature corresponding to scalar particles emission. We also obtain the radiation temperatures for the de Sitter, Rindler and Schwarzschild-de Sitter space-times. In a particular case when two horizons in Schwarzschild-de Sitter space-time coincides the Nariai temperature is recovered. The thermodynamical entropy of a black hole is calculated for Schwarzschild-de Sitter space-time having two horizons.

\end{abstract}

\section{Introduction}

Hawking investigated the radiation of scalar particles by black holes with the help of the Wick Rotation method \cite{Hawking} applied for the gravitational collapse. His result is based on semi-classical calculations and the thermal radiation arises quantum mechanically due to a horizon of the Schwarzschild space-time.
Thus, the field theory is quantized on the classical curved space-time background.
Tunnelling scalar particles via the spherically symmetric Schwarzschild black holes was studied in \cite{Hawking1}, \cite{Gibbons} but without affecting the background. Later the Hawking radiation was considered as a quantum tunnelling effect near the horizon and shown that it exhibits a thermal spectrum \cite{Wilczek}, \cite{Wilczek1}. Such an approach explores the WKB approximation to evaluate the tunnelling probability trajectory through the horizon that is classically forbidden. This method was successfully applied to study black hole radiation for different space-times such as Kerr \cite{Qing}, \cite{Zhang}, \cite{Mann}, the 3-dimensional BTZ black hole \cite{Zerbini}, \cite{Wu}, Vaidya black hole \cite{Jun}, Taub-NUT \cite{Mann1}, and G\"{o}del \cite{Mann2}. The Unruh temperature \cite{Unruh} is retrieved for a Rindler space-time \cite{Padmanabhan}, \cite{Mann}.
A black hole can radiate particles with different spins at some temperature so that all kind of particles can be produced. It was shown that spin-1/2 and spin-3/2 fermions and photons can be created by black holes having the same temperature as for the case of radiation of scalar particles \cite{Mann}, \cite{Mann3}, \cite{Jing}, \cite{Majhi}, \cite{Yale}. Authors \cite{Mann}, \cite{Mann3} have used the WKB approximation at leading order in Planck constant $\hbar$. Thus, the method of quantum tunnelling allows us to study thermodynamic properties of black holes and to understand the physical process of the black hole radiation.

To apply the tunnelling method one has to calculate the imaginary part of the action for the process of the emission through the horizon that is classically forbidden. Such imaginary part of the action is similar to the Boltzmann factor for the particle emission at the Hawking temperature. One of the approaches to evaluate the imaginary part of the action is the Null Geodesic Method developed in \cite{Wilczek}.
The second approach uses the Hamilton-Jacobi equation that is the generalization of the complex path analysis \cite{Padmanabhan}. These two methods are based on the estimation of the tunnelling probability, which is
the WKB approximation, for the classically forbidden trajectory via (from inside to outside) the horizon:
\begin{equation}
P=\exp\left(-2\mbox{Im} \frac{S}{\hbar}\right),
\label{1}
\end{equation}
where $S$ is the classical action of the path at the leading order in $\hbar$. For the second method, one should
explore the relativistic Hamilton-Jacobi equation to evaluate the imaginary part of the action of the emitted scalar particles. For the appropriate form of the action, one can solve the
relativistic Hamilton-Jacobi equation and to obtain the imaginary part of the action. Then with the help of Eq.(1) we can calculate the quantum tunnelling probability. Taking into account that a black hole possesses a definite temperature, we may obtain a temperature because Eq.(1) is similar to the Boltzmann expression for a thermal spectrum. It should be noted that a black hole can radiate particles with different spins and the emission spectrum of this radiation is similar to a black body radiation \cite{Page}. The tunnelling method was applied for the emission of scalar, spin-1/2, and spin-3/2 particles and photons. In this paper we use the tunnelling method to the case of the emission of massive spin-1 particles from non-rotating black holes.
Vector particles (spin-1 bosons) such as $Z$ and $W^\pm$ bosons are well-known and play very important role in Standard Model. Therefore, it is naturally to consider radiation of spin-1 bosons by black holes.
For simplicity we consider radiation of Proca fields in (1+1) dimensions. The probability of pair production of particles by the electric field (Schwinger's effect) depends on the spin of particles \cite{Kruglov}. We want to investigate the effect of the black hole radiation for massive spin-1 particles and to obtain the radiation temperature and entropy.
We apply a WKB approximation and the Hamilton-Jacobi ansatz \cite{Padmanabhan} to the Proca equation and show that particles are radiated by the black hole with the same Hawking temperature as in the case of scalar, spin-1/2, and spin-3/2 particles. Thus, the tunnelling approach shows that particles of spins $0$, $1/2$, $1$, and $3/2$ are emitted at the Hawking Temperature.
In our semi-classical calculations as well as in \cite{Mann1}, \cite{Mann2}, \cite{Mann3} the change of angular momentum of the black hole is neglected due to the emitting the spin particles. This is a good approximation for black holes with the mass much grater than the Planck mass.
We claim that massive spin-1 bosons are also emitted at the Hawking temperature.

We use the system of units: $c = G = k_B = 1$.

\section{Two dimensional space-times}

Let us consider the black hole in $(1+1)$ dimensions with the line element as follows:
\begin{equation}
ds^2=-A(r)dt^2+\frac{1}{B(r)}dr^2,
\label{2}
\end{equation}
with $c = 1$. Thus, we neglect the spherically symmetric 2-D part because only motion in radial direction is of our interest. The horizon in this space-time is in the point $r = r_0$ where $B(r)$ vanishes ($B(r_0)=0$). Near the horizon, we have $B(r)=B'(r_0)(r-r_0)+{\cal O}[(r-r_0)^2]$.
The Proca equations for vector particles are given by (see \cite{Kruglov2} where some aspects of the Proca equations are described)
\[
D_\mu\psi^{\nu\mu}+\frac{m^2}{\hbar^2}\psi^\nu=0,
\]
\vspace{-8mm}
\begin{equation}
\label{3}
\end{equation}
\vspace{-8mm}
\[
\psi_{\nu\mu}=D_\nu\psi_\mu-D_\mu\psi_\nu=\partial_\nu\psi_\mu-\partial_\mu\psi_\nu,
\]
where $D_\mu$ are covariant derivatives, and $\psi_\nu=(\psi_0,\psi_1)$. From Eqs.(3) one obtains the equation for the metric (2) as follows:
\begin{equation}
\sqrt{\frac{B(r)}{A(r)}}\partial_\mu\left(\sqrt{\frac{A(r)}{B(r)}}\psi^{\nu\mu}\right)+\frac{m^2}{\hbar^2}\psi^\nu=0,
\label{4}
\end{equation}
which is equivalent to the system of two second order equations
\[
\sqrt{A(r)B(r)}\partial_r\left[\sqrt{\frac{B(r)}{A(r)}}\left(\partial_t\psi_1-\partial_r\psi_0\right)\right]+\frac{m^2}{\hbar^2}\psi_0=0,
\]
\vspace{-8mm}
\begin{equation}
\label{5}
\end{equation}
\vspace{-8mm}
\[
\frac{1}{\sqrt{A(r)B(r)}}\partial_t\left[\sqrt{\frac{B(r)}{A(r)}}\left(\partial_t\psi_1-\partial_r\psi_0\right)\right]+\frac{m^2}{\hbar^2}\psi_1=0.
\]
With the help of the WKB approximation, we look for solutions to Eqs.(5) in the form
\begin{equation}
\psi_\nu=\left(c_0,c_1\right)\exp\left(\frac{i}{\hbar}S(t,r)\right),
\label{6}
\end{equation}
where the action is given by
\begin{equation}
S(t,r)=S_0(t,r)+\hbar S_1(t,r)+\hbar^2S_2(t,r)+....
\label{7}
\end{equation}
Replacing Eqs.(6),(7) into Eqs.(5), in the leading order to $\hbar$, one obtains equations as follows:
\[
B(r)\left[c_0(\partial_rS_0)^2-c_1(\partial_tS_0)(\partial_rS_0)\right]+m^2c_0=0,
\]
\vspace{-8mm}
\begin{equation}
\label{8}
\end{equation}
\vspace{-8mm}
\[
\frac{1}{A(r)}\left[c_0(\partial_tS_0)(\partial_rS_0)-c_1(\partial_tS_0)^2\right]+m^2c_1=0.
\]
The solution to Eqs.(8) can be obtained in the form
\begin{equation}
S_0=-Et+W(r)+C,
\label{9}
\end{equation}
where $E$ is an energy and $C$ is a (complex) constant. Replacing Eq.(9) into Eqs.(8) we obtain
the matrix equation
\begin{equation}
\left(\begin{array}{cc}
  B(W')^2+m^2 & BEW'  \\
  \frac{EW'}{A} & \frac{E^2}{A}-m^2
\end{array}\right)
\left(\begin{array}{c}
  c_0 \\
  c_1
\end{array}\right)=0,
\label{10}
\end{equation}
where $W'=\partial_rW$. Matrix equation (10) possesses the nontrivial solution if
the determinant of the matrix in Eq.(10) equals zero. Then we find
\begin{equation}
A(r)B(r)(W')^2=E^2-m^2A(r).
\label{11}
\end{equation}
From Eq.(11) one obtains
\begin{equation}
W_\pm(r)=\pm \int\sqrt{\frac{E^2-m^2A(r)}{A(r)B(r)}}dr.
\label{12}
\end{equation}
The signs in the square root are connected to the outgoing ($W_+$, $p_r=\partial_rS_0>0$) or ingoing ($W_-$, $p_r=\partial_rS_0<0$) motion of particles.
Here we consider a path in the direction $r$ from the inside to the outside of the horizon which corresponds to expression (12) with the sign $+$. The $W_\pm(r)$ has a simple pole at the horizon. In the regions $r>r_0$ and $r<r_0$ the integral in Eq.(12) is well defined and real, but if the path goes via the point $r_0$ the integral is not defined as $A^{-1}(r_0)=\infty$, $B^{-1}(r_0)=\infty$. To calculate the integral for crossing the horizon $r_0$, one can use a replacement $r_0\rightarrow r_0-i\varepsilon$ for outgoing particles \cite{Padmanabhan}. Thus, we specify the complex contour which can be used for the evaluating the integral around $r = r_0$.

The probability is normalized if Im$C$=-Im$W_-$ \cite{Mann1} and there is an absorbtion and no a reflection. An imaginary part of $C$ appears for Schwarzschild and other coordinates which are not well-defined across the horizon.
Thus, one needs for an imaginary part of $C$, that can be considered as boundary conditions, in our case of singular coordinate systems. It was mentioned in \cite{Hawking} that black holes may be created by classical collapse, but their disintegration is a quantum phenomenon. One can calculate the imaginary part of the integral (12) by the formula \cite{Bogolubov}
\begin{equation}
\frac{1}{r-i\varepsilon}=i\pi\delta(r)+{\cal P}\left(\frac{1}{r}\right),
\label{13}
\end{equation}
where ${\cal P}\left(\frac{1}{r}\right)$ is the principal value of $1/r$.
As a result, the probability of tunnelling a particle from inside to outside the horizon is
\begin{equation}
P=\exp\left(-\frac{4}{\hbar}\mbox{Im}W_+ \right)=\exp\left(-\frac{4\pi E}{\hbar\sqrt{\left(\partial_rA(r_0)\right)\partial_rB(r_0)}}+ {\cal O}(E^2)\right).
\label{14}
\end{equation}
The important step is to take into account that the classical absorption probability is unity. From Eq.(14) we obtain the temperature
\begin{equation}
T=\frac{\hbar\sqrt{\left(\partial_rA(r_0)\right)\partial_rB(r_0)}}{4\pi }.
\label{15}
\end{equation}
Thus, a temperature is associated with the horizon.
Expression (15) holds also for the black hole emission of particles with spins $0$, $1/2$, and $3/2$. The same result is valid also for spin-1 bosons in ($1+2$) dimensions \cite{Kruglov3}. The generalization to ($1+3$) dimensions is straightforward.

\subsection{Particular cases of space-times}

For the Schwarzchild black hole formed due to gravitational collapse of the matter, which is asymptotically flat,
$A(r)=B(r)=1-2M/r$ ($M$ is a mass of the black hole, $G=k_{\mbox B}=c=1$, and the event horizon is at $r_0=2M$).
In this case an observer at $l >2M$ detects the radiation at late times $t \rightarrow\infty$ compared with early times near the horizon $r_0 = 2M$. This radiation, obtained from Eqs.(14),(15), possesses Hawking's temperature $T_{\mbox{H}}=\hbar/(8\pi M)$ and we verify the Hawking thermal formula. For the de Sitter space-time $A(r)=B(r)=1-H^2r^2$ ($H=\dot{a}(t)/a(t)$ is the Hubble parameter, $a(t)$ is a scale factor, and the event horizon takes place at $r_0=1/H$), and from Eq.(15) we find the temperature $T=\hbar H/(2\pi)$. For the Rindler space-time $A(r)=B(r)=1+2gr$ ($g$ is an acceleration) and the temperature, obtained from Eqs.(14),(15), becomes the Unruh temperature $T=\hbar g/(2\pi)$.
The temperatures for these cases are determined by the horizons. Thus, the expression (15) can be used for evaluating the temperature for different metrics.

Let us consider the Schwarzschild-de Sitter space-time with the function
\begin{equation}
A(r)=B(r)=1-\frac{2M}{r}-\frac{r^2}{L^2}.
\label{16}
\end{equation}
Here $L=\sqrt{3/\Lambda}$ and $\Lambda$ is a cosmological constant. This space-time can possess two horizons corresponding to the solution of the equation $A(r)=B(r)=0$. We note that for $0<M/L<1/3\sqrt{3}$ there are three real roots with the link $r_1+r_2+r_3=0$. Therefore, only positive (two or one) roots correspond to event horizons. We find
\begin{equation}
1-\frac{2M}{r}-\frac{r^2}{L^2}=-\frac{(r-r_1)(r-r_2)(r-r_3)}{rL^2},
\label{17}
\end{equation}
where for $0<M/L<1/\sqrt{27}$ roots are given by
\begin{equation}
r_1=-\frac{2L}{\sqrt{3}}\cos\frac{\psi}{3},~~r_2=\frac{2L}{\sqrt{3}}\cos\frac{\pi+\psi}{3},
~~r_3=\frac{2L}{\sqrt{3}}\cos\frac{\pi-\psi}{3},
\label{18}
\end{equation}
and $\cos\psi=\sqrt{27}M/L$. Thus the root $r_1$ is negative and does not have physical meaning because $r\geq 0$, but $r_2$, $r_3$ ($r_2<r_3$) correspond to black hole event horizon and the cosmological horizon, respectively. We imply that particles cross two horizons. For the particle emissions the contour in Eq.(12) belongs the upper complex
plane. From Eqs.(16)-(18) we obtain
\begin{equation}
\frac{1}{|A(r)|}=\frac{rL}{2\sin\frac{\psi}{3}\left(r+\frac{2L}{\sqrt{3}}\cos\frac{\psi}{3}\right)}
\left(\frac{1}{r-r_3-i\varepsilon}-\frac{1}{r-r_2-i\varepsilon}\right).
\label{19}
\end{equation}
Then from Eqs.(12),(13),(19), one finds
\begin{equation}
\mbox{Im} W_+=\frac{E\pi L}{2\sqrt{3}\cos\frac{\psi}{3}}.
\label{20}
\end{equation}
From Eqs.(14),(20) and the Boltzmann expression we obtain the emission temperature for the general case of event and cosmological horizons
\begin{equation}
T=\frac{\hbar\sqrt{3}\cos\frac{\psi}{3}}{2\pi L}.
\label{21}
\end{equation}
It should be noted that simple expression (21) obtained is different from one given in \cite{Li}, \cite{Shankaranarayanan}.
\begin{figure}[h]
\includegraphics[height=4.0in,width=4.0in]{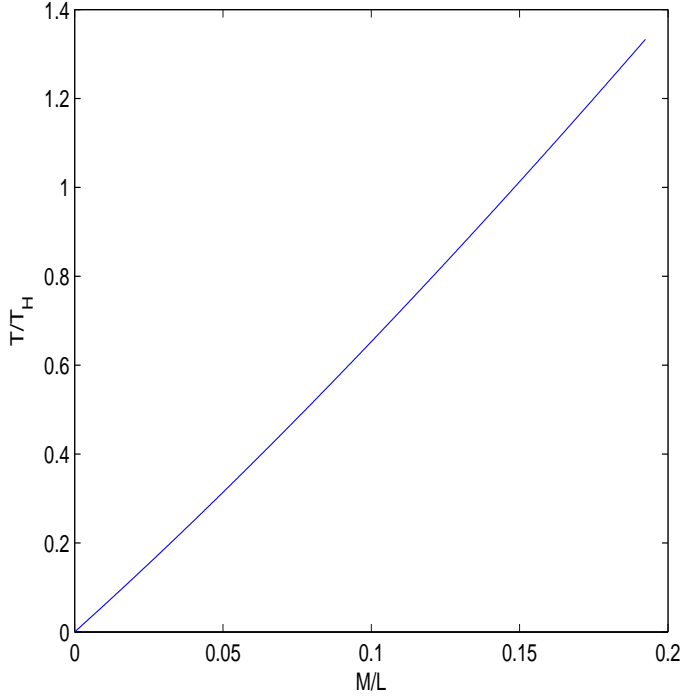}
\caption{\label{fig.1}The function $T/T_H$ versus $M/L$.}
\end{figure}
Also the temperature (21) corresponding to two horizons is not the sum of two temperatures associated with event and cosmological horizons. If two horizons coincides, $\psi=0$, $r_2=r_3$, and we recover the Nariai \cite{Nariai} temperature
\[
T_N=\frac{\hbar\sqrt{3}}{(2\pi L)}.
\]
The plot of the function $T/T_H$ versus $M/L$ is given in Fig.1.

\section{The thermodynamical entropy of a black hole for Schwarzschild-de Sitter space-time}

Now we calculate the thermodynamical entropy of a black hole when there are two horizons with unequal surface gravity. For a given temperature within classical thermodynamics one can use the change in the entropy as follows:
$dS(E) = dE/T(E)$. After integrating this equation we can calculate the function S(E) up to an additive constant. One can assume that for the Schwarzschild-de Sitter space-time an energy equals the mass of the black-hole ($c=1$) $E = M$.  Thus, to obtain the entropy of the black-hole we explore the relation
\begin{equation}
 dS = \frac{dM}{T(M)}.
\label{22}
\end{equation}
Integrating the equation (22) for Schwarzschild space-time leads to correct Hawking entropy ($\hbar=1$) $S_H=4\pi M^2$. Replacing the temperature (21) into Eq.(22), we obtain
\begin{equation}
 S = \frac{2\pi L}{\hbar\sqrt{3}}\int\frac{dM}{\cos\left(\frac{1}{3}\arccos\frac{\sqrt{27}M}{L}\right)}.
\label{23}
\end{equation}
After integration, we find entropy of a black hole for Schwarzschild-de Sitter space-time
\begin{equation}
 S = \frac{2\pi L^2}{\hbar 3}\left[-\ln\cos\left(\frac{1}{3}\arccos\frac{\sqrt{27}M}{L}\right)
 +2\cos^2\left(\frac{1}{3}\arccos\frac{\sqrt{27}M}{L}\right)\right]+C.
\label{24}
\end{equation}
An additive constant $C$ can be determined from an additional consideration.
The plot of the function $S\Lambda/2\pi$ versus $M/L$ ($L=\sqrt{3/\Lambda}$) for $C=0$ is given in Fig. 2.
For Nariai space-time, $\sqrt{27}M/L=1$, we obtain entropy
\begin{equation}
 S_N = \frac{4\pi L^2}{\hbar 3}+C.
\label{25}
\end{equation}
We suppose that the constant $C=0$.
\begin{figure}[h]
\includegraphics[height=4.0in,width=4.0in]{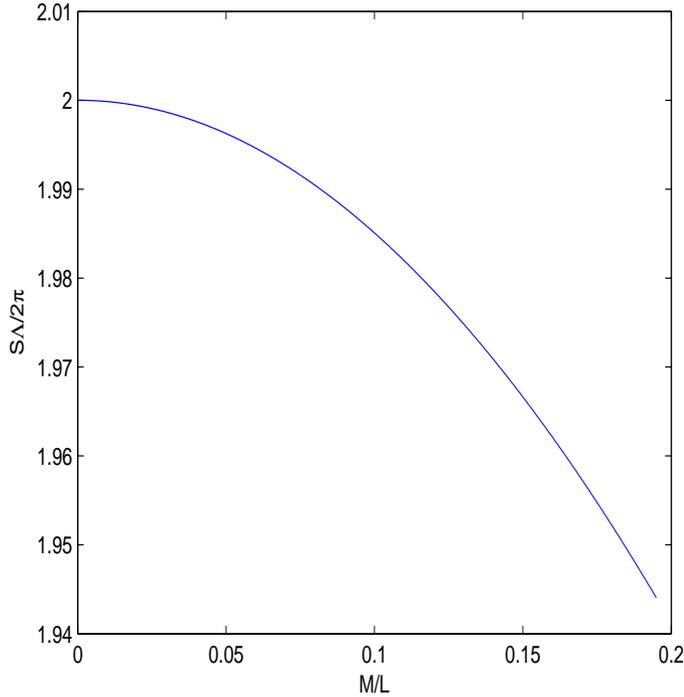}
\caption{\label{fig.2}The function $S\Lambda/2\pi$ versus $M/L$ (C=0).}
\end{figure}
It should be mentioned that there is a discussion of different expressions for temperature and entropy for Schwarzschild-de Sitter space-time in literature \cite{Padmanabhan}, \cite{Shankaranarayanan}. As a result, at present time, all expressions for temperature and entropy for Schwarzschild-de Sitter space-time are questionable. We imply that formulas for temperature (21) and entropy (24) for Schwarzschild-de Sitter space-time (at $C=0$) hold also for (3+1) dimensions for any spins.

\section{Discussion}

It was demonstrated with the help of the quantum tunnelling method that the Hawking temperature takes place for black holes emission of vector bosons due to the presence of the horizon. We have obtained the temperature and entropy for the Schwarzschild-de Sitter space-time with two horizons (event and cosmological), Eqs.(21), (24), that is our main result. The plots of the functions $T$ and $S$ versus the mass of black hole $M$ are given in figures 1, 2, respectively. For a particular case we found temperature and entropy of a black hole for Nariai space-time.

To calculate the radiation from evaporating black holes the Bogoliubov transformations can be used \cite{Birrel}. But the tunnelling method is simpler and can be applied for a variety of metrics for different spin particles.
It should be noted that we consider here only eternal black holes, which gives gravitational backgrounds.
Evaluating corrections to the tunnelling probability using the conservation of energy can solve the information paradox \cite{Bekenstein} which claims that information can be lost in a black hole, so that different physical states can present in the final state. But the information loss is in the contradiction with the conservation of information
and entropy. It should be noted that a thermal spectrum is inconsistent with the energy conservation because the background geometry is fixed.
One may compute corrections to the emission temperature by taking into consideration of higher orders in WKB. As a result, taking into account energy conservation \cite{Zhang1} during the emission of particles through tunnelling via the horizon and back-reaction effects \cite{Medved}, one obtains non-thermal corrections to the black-hole radiation spectrum. It should be noted that the $E^2$ correction to the Boltzmann expression arises from the energy conservation \cite{Wilczek1} and the spectrum is not precisely thermal. Non-thermality gives correlations in the radiation having the information. The correlations between Hawking radiations result in the amount of information contained in this correlation which equals to the information loss paradox \cite{Zhang1}. Thus, the statement that the information is lost during black hole evaporation is based on the assumption of thermality of the spectrum. Readers can find the discussion of the information paradox for evaporating black holes in \cite{Polchinski}, \cite{Hossenfelder}, \cite{Hawking2}.

\end{document}